\documentclass[aip,reprint]{revtex4-1}
\usepackage{graphicx}
\usepackage{xcolor}

\begin{document}
\onecolumngrid
\noindent
\textbf{PLEASE NOTE:} This article may be downloaded for personal use only. Any other use requires prior permission of the author and AIP Publishing. This article appeared in Rev. Sci. Instrum. \textbf{92}, 065109 (2021) and may be found at https://doi.org/10.1063/5.0047609
\vspace{2mm}

\title{Sapphire optical viewport for high pressure and temperature applications} 

\author{Till Ockenfels}
\email{ockenfels@iap.uni-bonn.de}
\author{Frank Vewinger}%
 
\author{Martin Weitz}
\affiliation{ 
Institut für Angewandte Physik, Universität Bonn, Wegelerstr. 8, D-53115 Bonn, Germany}


\begin{abstract}
We describe the design of a soldered sapphire optical viewport, useful for spectroscopic applications of samples at high temperatures and high pressures. The sapphire window is bonded via active soldering to a metal flange with a structure of two c-shaped rings made of different metallic materials in between, as to mitigate thermally induced stress. A spectroscopic cell equipped with two of the optical viewports has been successfully operated with alkali metals in a noble gas environment at temperatures in the range $20\,$°C to $450\,$°C at noble gas pressures from $10^{-6}\,$mbar to $330\,$bar. At the upper pressure range, we observe a leakage rate smaller than our readout accuracy of $30\,$mbar per day.
\end{abstract}


\maketitle 
\section{Introduction}
Optical spectroscopy is a powerful technique in fields ranging from atomic physics over chemistry and geology to industrial usage\cite{svanberg2004laser,khalafinejad2017exoplanetary,hammes2005spectroscopy,leenen2019rapid,pelletier1999analytical}. In several applications, the sample under investigation is a medium at high temperature or high pressure\cite{takeo1957broadening,christopoulos2018rubidium,vogler2018probing,Spearrin14,Brunsgaard}, leading to technical challenges in sealing the optical access to the medium’s container.

For optical access sapphire windows are frequently used as they withstand higher operation temperature and possess larger mechanical hardness compared to common glass materials. For high temperature spectroscopy at lower pressure, all-sapphire spectroscopic cells have been developed, making use of gluing the joints between sapphire parts\cite{sarkisyan2001sub}. Window-based spectroscopy cells viable also for high pressure operation are mostly built by joining sapphire optical windows with a metal cell body by seals made of organic components\cite{alberigs1974improved} or C-rings\cite{duignan1989metal}. However, organic sealings are not suited for experiments with aggressive substances as e.g. alkalis or high temperatures as they dismantle, and metallic C-ring seals become problematic at higher temperatures due to the different expansion coefficients of sapphire and sealing material. 

In comparison both a reduced leakage rate as well as an enhanced reliability is expected with the use of soldered metal-glass connections. Common metal glass joints, manufactured using active soldering, use relatively thin metal links attached to the glass. This can be understood from the required mechanical flexibility to sustain thermal stress occurring from the different thermal expansion coefficients during the soldering process\cite{chang2019active,o2005user}, imposing limits to high pressure operation.

Here we report a construction of a sapphire viewport where sealing is achieved by active soldering a sapphire optical window to a metal flange utilizing a stack of two intermediate rings, were the first ring is made of Kovar whose thermal expansion coefficient is close to the one of sapphire\cite{yim1974thermal,data_carpenter_tech_corp} and a second metal ring made of the softer metal copper to ensure mechanical flexibility. The construction has been tested to withstand both vacuum conditions and noble gas pressures (argon and helium)  of up to $330\,$bar in the cell at temperatures between $20\,$°C and $450\,$°C. Spectroscopic investigations of rubidium-buffer gas mixtures at the described experimental conditions were successfully performed using a cell equipped with the optical viewports.
\section{Construction}
\begin{figure}
\includegraphics[width=0.45\textwidth]{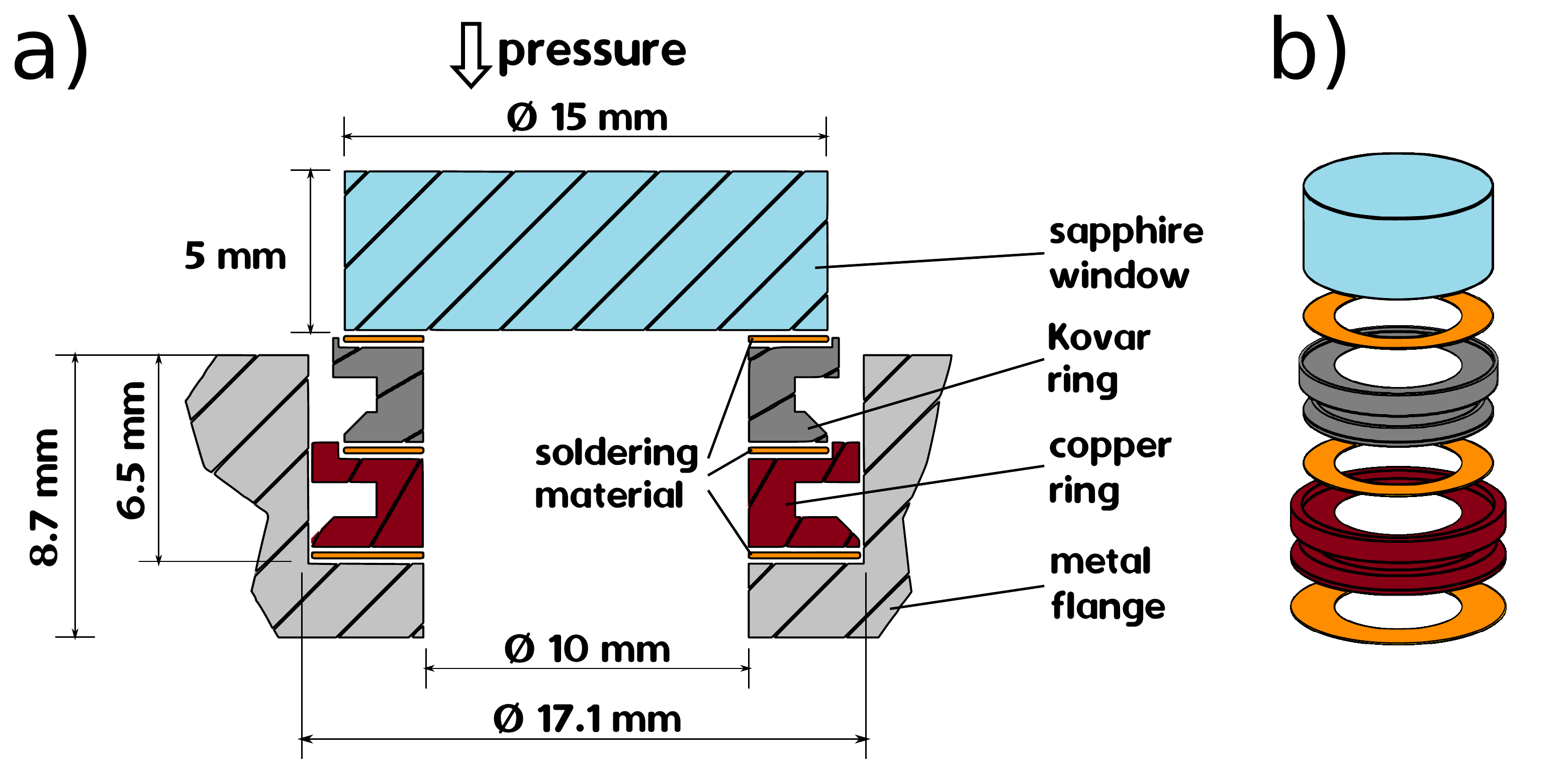}
\caption{\label{fig:drawing} a) Cross-section drawing of the soldered metal sapphire viewport. A circular sapphire window (top) is placed on the metal flange (bottom), with two metallic rings in between. A Kovar ring directly connected to the window is followed by a compensation ring made of copper. The different components are connected using an active soldering process, the soldering alloy is indicated by the orange rings. b) Schematic drawing showing the expanded assembly omiting the metal flange for better visibility.}
\end{figure}

A schematic of the viewport assembly is shown in Fig.~\ref{fig:drawing}.
A sapphire window (thermal expansion coefficient $\approx5.3\times10^{-6}\,$K$^{-1}$) is placed on a ring made of Kovar, a metallic nickel-cobalt alloy of very similar thermal expansion coefficient ($\approx5\times10^{-6}\,$K$^{-1}$) to prevent defects in the soldering metal-sapphire joint\cite{mishra2019recent}. This first ring is placed on top of a second “compensation” ring made of the soft metal copper, which itself is placed onto a stainless steel (type 1.4841) flange, which can then be connected to the metal body of the pressure cell. Both rings have grooves in the outer surface, leading to a c-shaped cross section, as to reduce the thickness of the material and therefore increasing mechanical flexibility and to mitigate mechanical stress acting on the sapphire crystal window while still providing a large contact area for the soldering connection with the adjacent parts. Given the desired high pressure operation of the viewport, the thickness of the rings must be chosen suitably to maintain sufficient mechanical stability.

\begin{figure}
\includegraphics[width=0.45\textwidth]{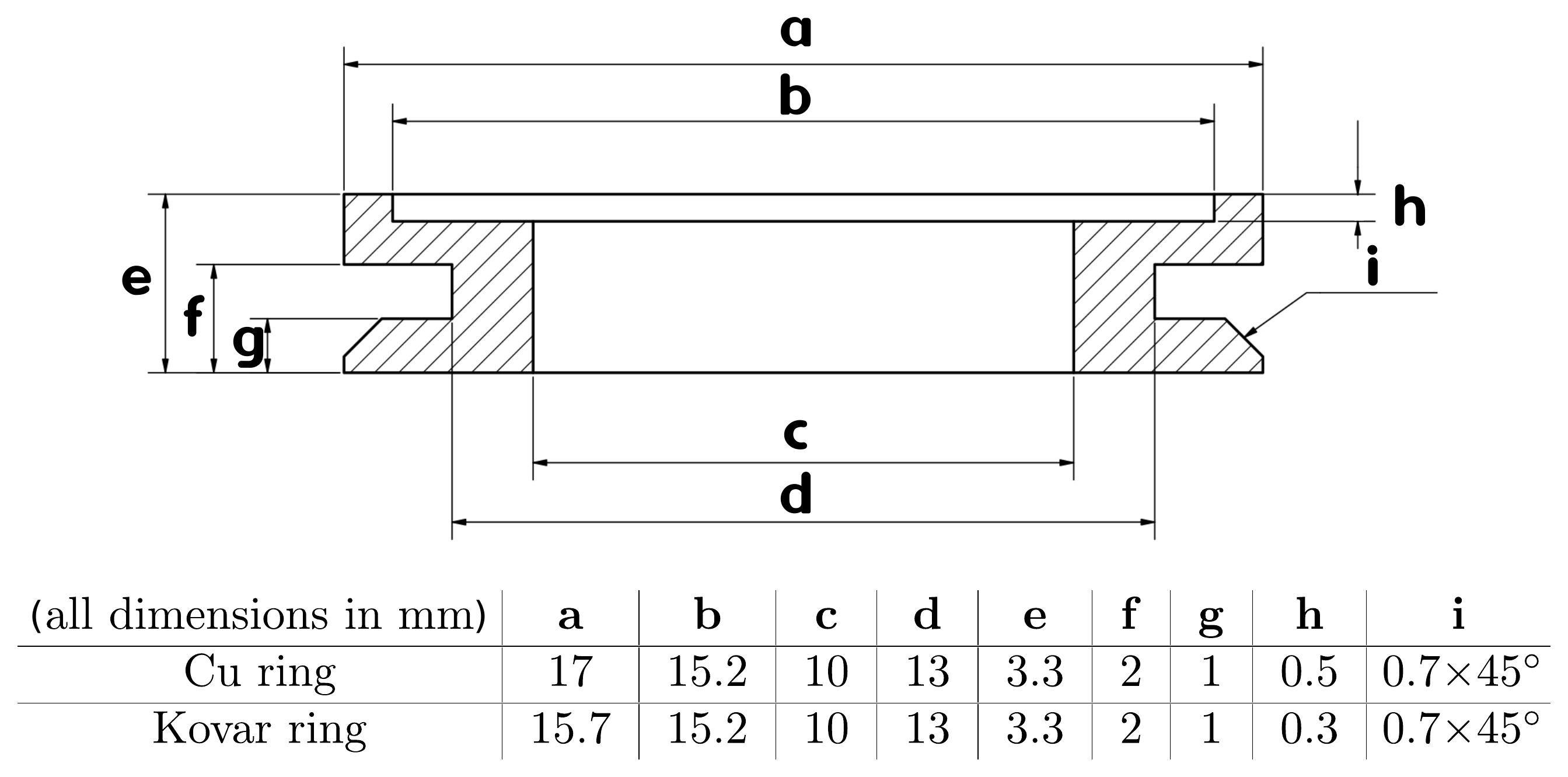}
\caption{\label{fig:dimensions} Cross-section drawing of the metallic rings with the associated dimensions given in the bottom table. The drawing is not to scale.}
\end{figure}

To bond our sapphire window with $15\,$mm outer diameter and 5mm thickness to the flange, we used a c-shaped Kovar ring that on the upper side has a shallow indentation to host the soldering alloy and the window. The second ring made of oxygen-free copper also has an indentation on the top side to fit the Kovar ring and the soldering alloy. The exact dimensions of both rings are given in Fig.~\ref{fig:dimensions}. 
To increase their flexibility both rings have a $1\,$mm wide grooves, leading to a minimum wall thickness of $1.5\,$mm, sufficient for the chosen pressure range of around to $300\,$bars. For the use at considerably higher cell operation pressures, the dimensions should be scaled to adequate wall thicknesses.
\section{Soldering}
\begin{figure}
\includegraphics[width=0.45\textwidth]{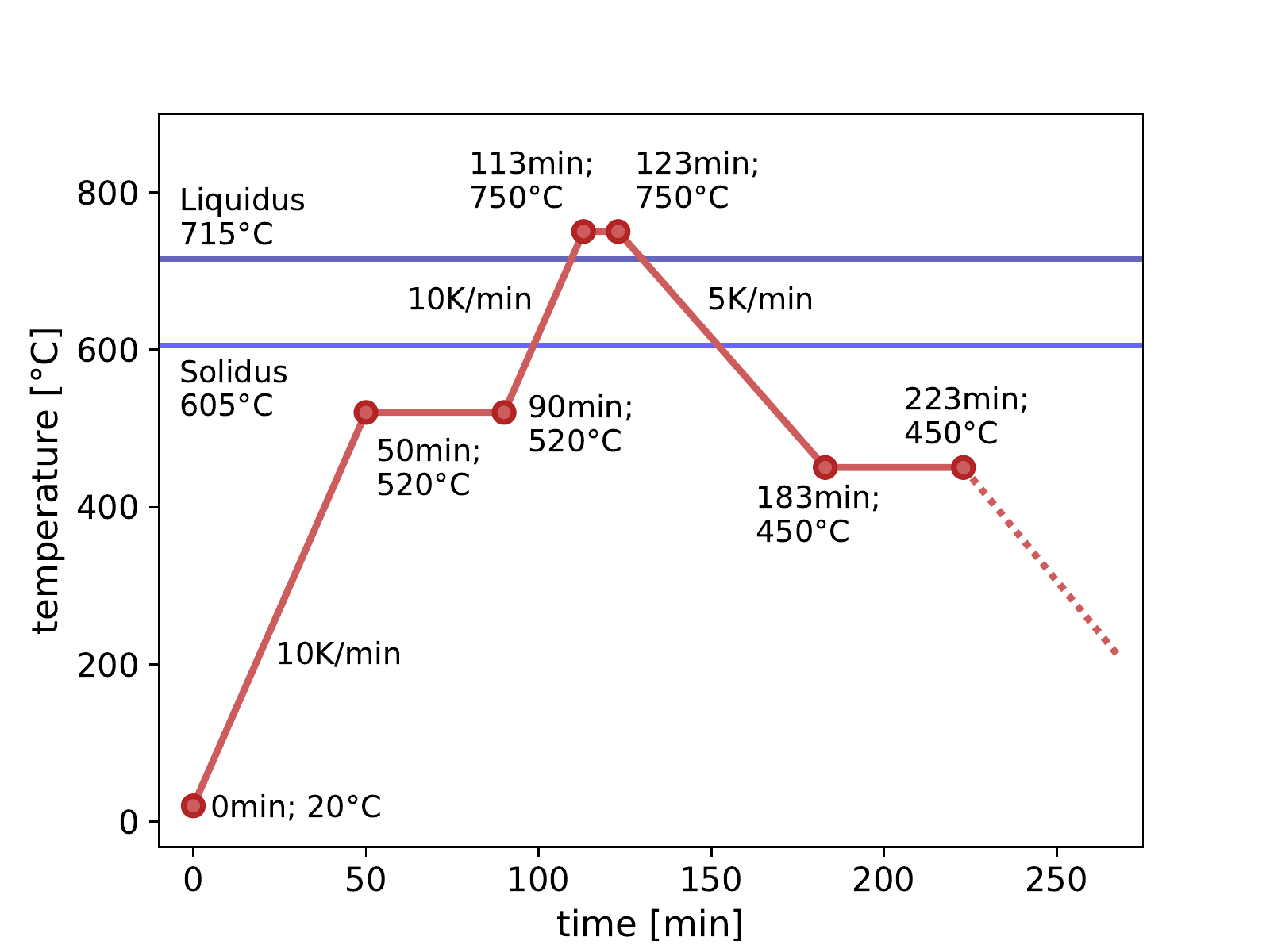}
\caption{\label{fig:plot} Temporal temperature profile used for soldering the viewport in a vacuum oven. Indicated by horizontal lines are the solidus and the liquidus temperatures of the used soldering alloy, respectively.}
\end{figure}

The different components of the viewport were joined at the interfaces by active soldering\cite{chang2019active}. The used soldering material (APA-7 from the supplier LOT-TEK with a composition of 59\% Ag, 27.25\% Cu, 12.5\% In, 1.25\% Ti) was laser-cut to the required dimensions from a foil of $0.1\,$mm thickness. Titanium is the active component of the soldering material which at high temperatures forms a chemical compound with oxygen from the sapphire material. In the bonding process no precedent metalization of the sapphire windows is needed, but the soldeing has to be done in a oxygen-free atmosphere\cite{chang2019active}. The melting point of the solder of around $700\,$°C is well above the operation temperatures in the envisaged application. The assembly is placed in an evacuated oven with an additional small weight on top of the window resulting in a force corresponding to $10\,$g per cm$^{2}$ at the contact area between window and solder to ensure a sufficient downforce in the soldering process. The used temperature curve shown in Fig.~\ref{fig:plot}
consists of a slow heating curve followed by a soldering stage and a subsequent cooling phase. To avoid thermal stress in the heating phase, we increase the temperature slowly, with a rate of $10\,$K$\,$min$^{-1}$ (which is small enough such that no large heat gradients between two sensors mounted on the components were measured), to a temperature of $750\,$°C, which is above the melting temperature of the soldering alloy. We keep the temperature constant at this value for about ten minutes to ensure a good connection between all parts in the soldering process. Subsequently, the temperature is reduced at a rate of $5\,$K$\,$min$^{-1}$ until room temperature is reached. To further reduce thermal stress on the construction we have implemented 40 minute-spans of constant temperature, which allow the whole assembly to fully thermalize. On the rising slope this is done at a temperature of $520\,$°C, on the falling slope at $450\,$°C. The temperature values for the holding times at which the temperature was left constant were chosen such that the solder is already soft and starts to flow, as to relief tension in the connection and leading to a more uniform joint.

In a preliminary stage of our experiment, viewports have been constructed without the copper compensation ring. Despite the close matching of the thermal expansion coefficients of the Kovar ring to that of the sapphire material, operation of these viewports occasionally resulted in breaking of the soldering links upon heating to operational temperatures. This was attributed to tension via direct mechanical connection of the Kovar ring with the flange from thermal expansion. Further experiments then used the construction described above, with the compensation ring. This has tremendously improved the reliability of the apparatus.
\section{Viewport properties}
With the described construction, we have manufactured several sapphire viewports that have been tested to withstand pressures of up to $330\,$bars at temperatures from $20\,$°C to $450\,$°C with no visible damage after usage, and multiple heating-cooling cycles could be performed without damage to the seals or windows. To test the tightness of the seal  we used a gas cell filled with $200\,$bar of helium at a constant temperature of $300\,$°C. Over the time span of one month the pressure dropped by less than the $\approx1\,$bar reading accuracy of the used manometer, giving an upper limit for the leakage rate of $30\,$mbars per day. After extensive bake-out for several days at a temperature of $450\,$°C, vacuum pressures at the lower end of the high vacuum regime ($10^{-8}\,$mbar to $10^{-9}\,$mbar) were reached while pumping the container with a turbo molecular pump, demonstrating the low vacuum leakage rate in that regime. In our main experiment (see e.g. [\onlinecite{christopoulos2018rubidium}]) the cell equipped with the assembled viewports has been successfully operated with rubidium vapor at $300\,$°C operation temperature ($\approx0.4\,$mbar vapor pressure) subject to typically $200\,$bar of noble buffer gas pressure, and allowed to observe pressure broadened spectra, without observable degradation of the all metal/sapphire construction from the chemically aggressive rubidium vapor after several weeks.

\section{Conclusion}
In conclusion, the construction of an optical cell viewport employing the joining of optical windows to a metal flange of the cell via active soldering has been described. The usage of only metallic parts (apart from the sapphire window) makes the system promising for spectroscopic studies of chemically aggressive materials at high pressure and/or high temperature conditions and process monitoring. 

\begin{acknowledgments}
We thank M. Brock, W. Graf, R. Langen and P. Hänisch for operating the soldering oven and technical advice. Financial support by the DFG (WE 1748-15, 581412 and SFB/TR 185, 277625399) is acknowledged. 
\end{acknowledgments}
\section*{Data Availability}
The data that support the findings of this study are available from the corresponding author upon reasonable request.
\onecolumngrid

\bibliography{viewport.bib}

\providecommand{\noopsort}[1]{}\providecommand{\singleletter}[1]{#1}%
\begin{thebibliography}{18}%
\makeatletter
\providecommand \@ifxundefined [1]{%
 \@ifx{#1\undefined}
}%
\providecommand \@ifnum [1]{%
 \ifnum #1\expandafter \@firstoftwo
 \else \expandafter \@secondoftwo
 \fi
}%
\providecommand \@ifx [1]{%
 \ifx #1\expandafter \@firstoftwo
 \else \expandafter \@secondoftwo
 \fi
}%
\providecommand \natexlab [1]{#1}%
\providecommand \enquote  [1]{``#1''}%
\providecommand \bibnamefont  [1]{#1}%
\providecommand \bibfnamefont [1]{#1}%
\providecommand \citenamefont [1]{#1}%
\providecommand \href@noop [0]{\@secondoftwo}%
\providecommand \href [0]{\begingroup \@sanitize@url \@href}%
\providecommand \@href[1]{\@@startlink{#1}\@@href}%
\providecommand \@@href[1]{\endgroup#1\@@endlink}%
\providecommand \@sanitize@url [0]{\catcode `\\12\catcode `\$12\catcode
  `\&12\catcode `\#12\catcode `\^12\catcode `\_12\catcode `\%12\relax}%
\providecommand \@@startlink[1]{}%
\providecommand \@@endlink[0]{}%
\providecommand \url  [0]{\begingroup\@sanitize@url \@url }%
\providecommand \@url [1]{\endgroup\@href {#1}{\urlprefix }}%
\providecommand \urlprefix  [0]{URL }%
\providecommand \Eprint [0]{\href }%
\providecommand \doibase [0]{http://dx.doi.org/}%
\providecommand \selectlanguage [0]{\@gobble}%
\providecommand \bibinfo  [0]{\@secondoftwo}%
\providecommand \bibfield  [0]{\@secondoftwo}%
\providecommand \translation [1]{[#1]}%
\providecommand \BibitemOpen [0]{}%
\providecommand \bibitemStop [0]{}%
\providecommand \bibitemNoStop [0]{.\EOS\space}%
\providecommand \EOS [0]{\spacefactor3000\relax}%
\providecommand \BibitemShut  [1]{\csname bibitem#1\endcsname}%
\let\auto@bib@innerbib\@empty
\bibitem [{\citenamefont {Svanberg}(2004)}]{svanberg2004laser}%
  \BibitemOpen
  \bibfield  {author} {\bibinfo {author} {\bibfnamefont {S.}~\bibnamefont
  {Svanberg}},\ }\bibfield  {title} {\enquote {\bibinfo {title}
  {Laser-spectroscopic applications},}\ }in\ \href@noop {} {\emph {\bibinfo
  {booktitle} {Atomic and Molecular Spectroscopy}}}\ (\bibinfo  {publisher}
  {Springer},\ \bibinfo {year} {2004})\ pp.\ \bibinfo {pages}
  {389--460}\BibitemShut {NoStop}%
\bibitem [{\citenamefont {Khalafinejad}\ \emph {et~al.}(2017)\citenamefont
  {Khalafinejad}, \citenamefont {von Essen}, \citenamefont {Hoeijmakers},
  \citenamefont {Zhou}, \citenamefont {Klocov{\'a}}, \citenamefont {Schmitt},
  \citenamefont {Dreizler}, \citenamefont {Lopez-Morales}, \citenamefont
  {Husser}, \citenamefont {Schmidt} \emph
  {et~al.}}]{khalafinejad2017exoplanetary}%
  \BibitemOpen
  \bibfield  {author} {\bibinfo {author} {\bibfnamefont {S.}~\bibnamefont
  {Khalafinejad}}, \bibinfo {author} {\bibfnamefont {C.}~\bibnamefont {von
  Essen}}, \bibinfo {author} {\bibfnamefont {H.}~\bibnamefont {Hoeijmakers}},
  \bibinfo {author} {\bibfnamefont {G.}~\bibnamefont {Zhou}}, \bibinfo {author}
  {\bibfnamefont {T.}~\bibnamefont {Klocov{\'a}}}, \bibinfo {author}
  {\bibfnamefont {J.}~\bibnamefont {Schmitt}}, \bibinfo {author} {\bibfnamefont
  {S.}~\bibnamefont {Dreizler}}, \bibinfo {author} {\bibfnamefont
  {M.}~\bibnamefont {Lopez-Morales}}, \bibinfo {author} {\bibfnamefont {T.-O.}\
  \bibnamefont {Husser}}, \bibinfo {author} {\bibfnamefont {T.}~\bibnamefont
  {Schmidt}},  \emph {et~al.},\ }\bibfield  {title} {\enquote {\bibinfo {title}
  {Exoplanetary atmospheric sodium revealed by orbital motion-narrow-band
  transmission spectroscopy of hd 189733b with uves},}\ }\href@noop {}
  {\bibfield  {journal} {\bibinfo  {journal} {Astronomy \& Astrophysics}\
  }\textbf {\bibinfo {volume} {598}},\ \bibinfo {pages} {A131} (\bibinfo {year}
  {2017})}\BibitemShut {NoStop}%
\bibitem [{\citenamefont {Hammes}(2005)}]{hammes2005spectroscopy}%
  \BibitemOpen
  \bibfield  {author} {\bibinfo {author} {\bibfnamefont {G.~G.}\ \bibnamefont
  {Hammes}},\ }\href@noop {} {\emph {\bibinfo {title} {Spectroscopy for the
  biological sciences}}}\ (\bibinfo  {publisher} {John Wiley \& Sons},\
  \bibinfo {year} {2005})\BibitemShut {NoStop}%
\bibitem [{\citenamefont {Leenen}\ \emph {et~al.}(2019)\citenamefont {Leenen},
  \citenamefont {Welp}, \citenamefont {Gebbers},\ and\ \citenamefont
  {P{\"a}tzold}}]{leenen2019rapid}%
  \BibitemOpen
  \bibfield  {author} {\bibinfo {author} {\bibfnamefont {M.}~\bibnamefont
  {Leenen}}, \bibinfo {author} {\bibfnamefont {G.}~\bibnamefont {Welp}},
  \bibinfo {author} {\bibfnamefont {R.}~\bibnamefont {Gebbers}}, \ and\
  \bibinfo {author} {\bibfnamefont {S.}~\bibnamefont {P{\"a}tzold}},\
  }\bibfield  {title} {\enquote {\bibinfo {title} {Rapid determination of lime
  requirement by mid-infrared spectroscopy: A promising approach for precision
  agriculture},}\ }\href@noop {} {\bibfield  {journal} {\bibinfo  {journal}
  {Journal of Plant Nutrition and Soil Science}\ }\textbf {\bibinfo {volume}
  {182}},\ \bibinfo {pages} {953--963} (\bibinfo {year} {2019})}\BibitemShut
  {NoStop}%
\bibitem [{\citenamefont {Pelletier}\ \emph {et~al.}(1999)\citenamefont
  {Pelletier} \emph {et~al.}}]{pelletier1999analytical}%
  \BibitemOpen
  \bibfield  {author} {\bibinfo {author} {\bibfnamefont {M.~J.}\ \bibnamefont
  {Pelletier}} \emph {et~al.},\ }\href@noop {} {\emph {\bibinfo {title}
  {Analytical applications of Raman spectroscopy}}},\ Vol.\ \bibinfo {volume}
  {427}\ (\bibinfo  {publisher} {Blackwell science Oxford},\ \bibinfo {year}
  {1999})\BibitemShut {NoStop}%
\bibitem [{\citenamefont {Takeo}\ \emph {et~al.}(1957)\citenamefont {Takeo}
  \emph {et~al.}}]{takeo1957broadening}%
  \BibitemOpen
  \bibfield  {author} {\bibinfo {author} {\bibfnamefont {M.}~\bibnamefont
  {Takeo}} \emph {et~al.},\ }\bibfield  {title} {\enquote {\bibinfo {title}
  {Broadening and shift of spectral lines due to the presence of foreign
  gases},}\ }\href@noop {} {\bibfield  {journal} {\bibinfo  {journal} {Reviews
  of Modern Physics}\ }\textbf {\bibinfo {volume} {29}},\ \bibinfo {pages} {20}
  (\bibinfo {year} {1957})}\BibitemShut {NoStop}%
\bibitem [{\citenamefont {Christopoulos}\ \emph {et~al.}(2018)\citenamefont
  {Christopoulos}, \citenamefont {Moroshkin}, \citenamefont {Weller},
  \citenamefont {Gerwers}, \citenamefont {Forge}, \citenamefont {Ockenfels},
  \citenamefont {Vewinger},\ and\ \citenamefont
  {Weitz}}]{christopoulos2018rubidium}%
  \BibitemOpen
  \bibfield  {author} {\bibinfo {author} {\bibfnamefont {S.}~\bibnamefont
  {Christopoulos}}, \bibinfo {author} {\bibfnamefont {P.}~\bibnamefont
  {Moroshkin}}, \bibinfo {author} {\bibfnamefont {L.}~\bibnamefont {Weller}},
  \bibinfo {author} {\bibfnamefont {B.}~\bibnamefont {Gerwers}}, \bibinfo
  {author} {\bibfnamefont {R.}~\bibnamefont {Forge}}, \bibinfo {author}
  {\bibfnamefont {T.}~\bibnamefont {Ockenfels}}, \bibinfo {author}
  {\bibfnamefont {F.}~\bibnamefont {Vewinger}}, \ and\ \bibinfo {author}
  {\bibfnamefont {M.}~\bibnamefont {Weitz}},\ }\bibfield  {title} {\enquote
  {\bibinfo {title} {Rubidium spectroscopy at high-pressure buffer gas
  conditions: detailed balance in the optical interaction of an absorber
  coupled to a reservoir},}\ }\href@noop {} {\bibfield  {journal} {\bibinfo
  {journal} {Physica Scripta}\ }\textbf {\bibinfo {volume} {93}},\ \bibinfo
  {pages} {124006} (\bibinfo {year} {2018})}\BibitemShut {NoStop}%
\bibitem [{\citenamefont {Vogler}\ and\ \citenamefont
  {V{\"o}hringer}(2018)}]{vogler2018probing}%
  \BibitemOpen
  \bibfield  {author} {\bibinfo {author} {\bibfnamefont {T.}~\bibnamefont
  {Vogler}}\ and\ \bibinfo {author} {\bibfnamefont {P.}~\bibnamefont
  {V{\"o}hringer}},\ }\bibfield  {title} {\enquote {\bibinfo {title} {Probing
  the band gap of liquid ammonia with femtosecond multiphoton ionization
  spectroscopy},}\ }\href@noop {} {\bibfield  {journal} {\bibinfo  {journal}
  {Physical Chemistry Chemical Physics}\ }\textbf {\bibinfo {volume} {20}},\
  \bibinfo {pages} {25657--25665} (\bibinfo {year} {2018})}\BibitemShut
  {NoStop}%
\bibitem [{\citenamefont {Spearrin}\ \emph {et~al.}(2014)\citenamefont
  {Spearrin}, \citenamefont {Goldenstein}, \citenamefont {Jeffries},\ and\
  \citenamefont {Hanson}}]{Spearrin14}%
  \BibitemOpen
  \bibfield  {author} {\bibinfo {author} {\bibfnamefont {R.~M.}\ \bibnamefont
  {Spearrin}}, \bibinfo {author} {\bibfnamefont {C.~S.}\ \bibnamefont
  {Goldenstein}}, \bibinfo {author} {\bibfnamefont {J.~B.}\ \bibnamefont
  {Jeffries}}, \ and\ \bibinfo {author} {\bibfnamefont {R.~K.}\ \bibnamefont
  {Hanson}},\ }\bibfield  {title} {\enquote {\bibinfo {title} {Quantum cascade
  laser absorption sensor for carbon monoxide in high-pressure gases using
  wavelength modulation spectroscopy},}\ }\href {\doibase 10.1364/AO.53.001938}
  {\bibfield  {journal} {\bibinfo  {journal} {Appl. Opt.}\ }\textbf {\bibinfo
  {volume} {53}},\ \bibinfo {pages} {1938--1946} (\bibinfo {year}
  {2014})}\BibitemShut {NoStop}%
\bibitem [{\citenamefont {Hansen}, \citenamefont {Berg},\ and\ \citenamefont
  {Stenby}(2001)}]{Brunsgaard}%
  \BibitemOpen
  \bibfield  {author} {\bibinfo {author} {\bibfnamefont {S.~B.}\ \bibnamefont
  {Hansen}}, \bibinfo {author} {\bibfnamefont {R.~W.}\ \bibnamefont {Berg}}, \
  and\ \bibinfo {author} {\bibfnamefont {E.~H.}\ \bibnamefont {Stenby}},\
  }\bibfield  {title} {\enquote {\bibinfo {title} {High-pressure measuring cell
  for raman spectroscopic studies of natural gas},}\ }\href {\doibase
  10.1366/0003702011951434} {\bibfield  {journal} {\bibinfo  {journal} {Applied
  Spectroscopy}\ }\textbf {\bibinfo {volume} {55}},\ \bibinfo {pages} {55--60}
  (\bibinfo {year} {2001})},\ \Eprint
  {http://arxiv.org/abs/https://doi.org/10.1366/0003702011951434}
  {https://doi.org/10.1366/0003702011951434} \BibitemShut {NoStop}%
\bibitem [{\citenamefont {Sarkisyan}\ \emph {et~al.}(2001)\citenamefont
  {Sarkisyan}, \citenamefont {Bloch}, \citenamefont {Papoyan},\ and\
  \citenamefont {Ducloy}}]{sarkisyan2001sub}%
  \BibitemOpen
  \bibfield  {author} {\bibinfo {author} {\bibfnamefont {D.}~\bibnamefont
  {Sarkisyan}}, \bibinfo {author} {\bibfnamefont {D.}~\bibnamefont {Bloch}},
  \bibinfo {author} {\bibfnamefont {A.}~\bibnamefont {Papoyan}}, \ and\
  \bibinfo {author} {\bibfnamefont {M.}~\bibnamefont {Ducloy}},\ }\bibfield
  {title} {\enquote {\bibinfo {title} {Sub-{D}oppler spectroscopy by sub-micron
  thin {C}s vapour layer},}\ }\href@noop {} {\bibfield  {journal} {\bibinfo
  {journal} {Optics Communications}\ }\textbf {\bibinfo {volume} {200}},\
  \bibinfo {pages} {201--208} (\bibinfo {year} {2001})}\BibitemShut {NoStop}%
\bibitem [{\citenamefont {Alberigs}\ and\ \citenamefont
  {Penninger}(1974)}]{alberigs1974improved}%
  \BibitemOpen
  \bibfield  {author} {\bibinfo {author} {\bibfnamefont {J.}~\bibnamefont
  {Alberigs}}\ and\ \bibinfo {author} {\bibfnamefont {J.}~\bibnamefont
  {Penninger}},\ }\bibfield  {title} {\enquote {\bibinfo {title} {An improved
  window seal for high temperature-pressure spectroscopic flow cells},}\
  }\href@noop {} {\bibfield  {journal} {\bibinfo  {journal} {Review of
  Scientific Instruments}\ }\textbf {\bibinfo {volume} {45}},\ \bibinfo {pages}
  {460--461} (\bibinfo {year} {1974})}\BibitemShut {NoStop}%
\bibitem [{\citenamefont {Duignan}, \citenamefont {Gerhardt},\ and\
  \citenamefont {Whitney}(1989)}]{duignan1989metal}%
  \BibitemOpen
  \bibfield  {author} {\bibinfo {author} {\bibfnamefont {M.~T.}\ \bibnamefont
  {Duignan}}, \bibinfo {author} {\bibfnamefont {D.~J.}\ \bibnamefont
  {Gerhardt}}, \ and\ \bibinfo {author} {\bibfnamefont {W.}~\bibnamefont
  {Whitney}},\ }\bibfield  {title} {\enquote {\bibinfo {title} {Metal c ring
  window seal},}\ }\href@noop {} {\bibfield  {journal} {\bibinfo  {journal}
  {Review of Scientific Instruments}\ }\textbf {\bibinfo {volume} {60}},\
  \bibinfo {pages} {3537--3539} (\bibinfo {year} {1989})}\BibitemShut {NoStop}%
\bibitem [{\citenamefont {Chang}, \citenamefont {Huang},\ and\ \citenamefont
  {Tsao}(2019)}]{chang2019active}%
  \BibitemOpen
  \bibfield  {author} {\bibinfo {author} {\bibfnamefont {S.-Y.}\ \bibnamefont
  {Chang}}, \bibinfo {author} {\bibfnamefont {Y.-H.}\ \bibnamefont {Huang}}, \
  and\ \bibinfo {author} {\bibfnamefont {L.-C.}\ \bibnamefont {Tsao}},\
  }\bibfield  {title} {\enquote {\bibinfo {title} {Active solders and active
  soldering},}\ }in\ \href@noop {} {\emph {\bibinfo {booktitle}
  {Fillers-Synthesis, Characterization and Industrial Application}}}\ (\bibinfo
   {publisher} {IntechOpen},\ \bibinfo {year} {2019})\BibitemShut {NoStop}%
\bibitem [{\citenamefont {O'Hanlon}(2005)}]{o2005user}%
  \BibitemOpen
  \bibfield  {author} {\bibinfo {author} {\bibfnamefont {J.~F.}\ \bibnamefont
  {O'Hanlon}},\ }\href@noop {} {\emph {\bibinfo {title} {A user's guide to
  vacuum technology}}}\ (\bibinfo  {publisher} {John Wiley \& Sons},\ \bibinfo
  {year} {2005})\BibitemShut {NoStop}%
\bibitem [{\citenamefont {Yim}\ and\ \citenamefont
  {Paff}(1974)}]{yim1974thermal}%
  \BibitemOpen
  \bibfield  {author} {\bibinfo {author} {\bibfnamefont {W.}~\bibnamefont
  {Yim}}\ and\ \bibinfo {author} {\bibfnamefont {R.}~\bibnamefont {Paff}},\
  }\bibfield  {title} {\enquote {\bibinfo {title} {Thermal expansion of
  {A}l{N}, sapphire, and silicon},}\ }\href@noop {} {\bibfield  {journal}
  {\bibinfo  {journal} {Journal of Applied Physics}\ }\textbf {\bibinfo
  {volume} {45}},\ \bibinfo {pages} {1456--1457} (\bibinfo {year}
  {1974})}\BibitemShut {NoStop}%
\bibitem [{dat(1990)}]{data_carpenter_tech_corp}%
  \BibitemOpen
  \href@noop {} {\emph {\bibinfo {title} {Technical Datasheet CarTech®Kovar®
  Alloy}}},\ \bibinfo {organization} {CRS Holdings Inc.} (\bibinfo {year}
  {1990}),\ \bibinfo {note} {cRS Holdings Inc. is a subsidiary of Carpenter
  Technology Corporation}\BibitemShut {NoStop}%
\bibitem [{\citenamefont {Mishra}\ \emph {et~al.}(2020)\citenamefont {Mishra},
  \citenamefont {Sharma}, \citenamefont {Jung},\ and\ \citenamefont
  {Jung}}]{mishra2019recent}%
  \BibitemOpen
  \bibfield  {author} {\bibinfo {author} {\bibfnamefont {S.}~\bibnamefont
  {Mishra}}, \bibinfo {author} {\bibfnamefont {A.}~\bibnamefont {Sharma}},
  \bibinfo {author} {\bibfnamefont {D.}~\bibnamefont {Jung}}, \ and\ \bibinfo
  {author} {\bibfnamefont {J.}~\bibnamefont {Jung}},\ }\bibfield  {title}
  {\enquote {\bibinfo {title} {Recent advances in active metal brazing of
  ceramics and process},}\ }\href@noop {} {\bibfield  {journal} {\bibinfo
  {journal} {Metals and Materials International}\ }\textbf {\bibinfo {volume}
  {26}},\ \bibinfo {pages} {1087--1098} (\bibinfo {year} {2020})}\BibitemShut
  {NoStop}%
\end{thebibliography}%

\end{document}